\documentclass[12pt]{article}
\usepackage{amsmath}
\usepackage{amssymb}
\usepackage{amsfonts} 
\newcommand{\be}{\begin{equation}}
\newcommand{\ba}{\begin{eqnarray}}
\newcommand{\ee}{\end{equation}}
\newcommand{\ea}{\end{eqnarray}}
\newcommand{\cosech} { {\rm cosech}}
\DeclareMathOperator{\sech}{sech}

\newcommand{\ket}[1]{\ensuremath{\left|#1\right\rangle}}

\begin{document}

\title{Group theoretic approach to rationally extended shape invariant potentials}

\author{Rajesh Kumar Yadav$^{a}$\footnote{e-mail address: rajeshastrophysics@gmail.com (R.K.Y)}, Nisha Kumari$^{a}$\footnote{e-mail address: nishaism0086@gmail.com (N.K)}, Avinash Khare$^{b}$\footnote {e-mail address: khare@iiserpune.ac.in (A.K)} and \\
 Bhabani Prasad Mandal$^{a}$\footnote{e-mail address: bhabani.mandal@gmail.com (B.P.M).}}
 \maketitle
{$~^a$ Department of Physics, Banaras Hindu University, Varanasi-221005, INDIA.\\
$~^b$Raja Ramanna Fellow, Indian Institute of Science Education and Research (IISER), Pune-411021, INDIA.}

\begin{abstract}

The exact bound state spectrum of rationally extended shape invariant real as well as $PT$ symmetric 
complex potentials are obtained by using potential group approach. The generators of the potential 
groups are modified by introducing a new operator $U(x,J_3\pm\frac{1}{2})$ to express the Hamiltonian 
corresponding to these extended potentials in terms of Casimir operators. 
Connection between the potential algebra and the shape invariance is elucidated.

\end{abstract}

\section{Introduction}

 The ideas of supersymmetric quantum mechanics (SQM) and shape 
invariant (SI) potentials \cite{cks} have played useful role in discovering
new exactly solvable potentials.
 Recently after the discovery of exceptional orthogonal polynomials (EOPs) (also known as $X_m$ 
Laguerre and $X_m$ Jacobi polynomials \cite{dnr1,dnr2,xm2}, the SQM 
and SI ideas have been used to discover new 
shape invariant potentials with translation. Unlike the usual orthogonal polynomials (which starts with degree $m=0$), 
these EOPs start with degree $m \geq 1$ and still form a complete orthonormal set with respect to a
 positive definite inner product defined over a compact interval. The properties of these $X_m$ exceptional 
orthogonal polynomials have been studied in detail in Ref. \cite{hos,hs,gom,qu,dim}. After the discovery of these 
 two polynomials the new rationally extended potentials have been
 obtained whose solutions are in terms of $X_m$ EOPs \cite{os}.
 In particular for $m=0$ these corresponds to the conventional shape invariant potentials given in Ref. \cite{cks} and 
for $m=1$ the obtained potentials are the rationally extended translationally shape invariant (w.r.t potential 
parameters) potentials whose solutions are in terms of $X_1$ Laguerre or $X_1$ Jacobi polynomials \cite{que,bqr}.
 Later on, the rational extension of the other exactly solvable potentials has 
also been considered, but the solutions of these potentials are
 not in the forms of EOPs, rather they are in the form of some new types of polynomials \cite{op4}. 
In these potentials the usual SI property is no more valid, rather they exhibit an unfamiliar extended SI 
property in which the partner potential is obtained by translating both the potential parameter $A$ 
(as in the conventional case) and $m$, the degree of the polynomial arising in the denominator. 
All of these rationally extended potentials are isospectral to their conventional counterparts.    
  
 Apart from the SQM and SI approach as mentioned above, the above extended 
potentials and their bound states 
 have been also obtained through different approaches such as Point canonical 
transformation (PCT) \cite{pct,que}
 approach, Darboux Crum transformation \cite{dt,dnr2,xm2,rsa}, Darboux-Backlund transformation \cite{yg1,yg2,yg3} 
and Prepotential approach \cite{clh1,clh2}. The multi-indexed extension of 
some of these potentials have also been done by 
using multi-step Darboux-Backlund transformation and higher-order SQM \cite{mdbt,hsusy}.
EOPs have further been studied in different quantum mechanical systems such as quantum Hamilton-Jacobi
formalism \cite{hj}, position dependent mass system \cite{pdm}, N-fold SUSY \cite{nfold1,nfold2} ,Fokker-Planck equation \cite{fplank}, conditionally 
exactly solvable potentials \cite{codext1,codext2}, quantum mechanical scattering \cite{qscat1,qscat2,qscat3} and time dependent potentials \cite{tdse}.

In last few years the idea of PT-symmetry has attracted lot of attention.
Consistent quantum theories with $PT$ symmetric complex potentials have been developed exclusively over 
the past one and half decades \cite{bender}. After the discovery of EOPs, 
the $PT$ symmetric complex potentials have also been extended rationally 
\cite{bqr}. The solutions of these rationally extended 
$PT$ symmetric complex potentials are also written in terms of EOPs. 

In addition to the above approaches, there is an independent powerful approach i.e., potential algebra (or group theoretical) approach \cite{potalg1,potalg2,potalg3,potalg4,potalg5,potalg6,potalg7,potalg8}, 
by which one can obtain the exact spectrum of some of these exactly solvable potentials. 
Alhassid et.al. applied  this approach earlier to several solvable potentials 
and obtained 
the exact spectrum. Later on the connection of this approach with the SI has also been 
established \cite{sipotalg1}.

In the present work we extend the ideas of Alhassid et al to the case of 
rationally extended potentials and obtain the   
exact bound state spectrum using potential algebra approach.
For the concreteness we consider 
two examples: (i) extended generalized P\"oschl-Teller (GPT) real potential and (ii) extended $PT$ symmetric 
complex Scarf-II potential. The potential group $SO(2,1)$ and $sl(2,\mathbb{C})$ are used for this purpose 
for real and $PT$ symmetric complex potential respectively. By modifying the generators $(J_{\pm})$ of 
these groups by introducing a new operator $U(x,J_3\pm\frac{1}{2})$, the Hamiltonian corresponding to
these rationally extended systems are written in terms of Casimir invariants of the relevant group.
The exact bound state spectrum of these SI potentials are obtained in a closed form.
In this work we show that 
rationally extended real as well as $PT$ symmetric complex potentials whose solutions
are written in terms of EOPs are solved using potential group algebra in an elegant manner.
We first consider the case where solutions are in terms of $X_1$ EOPs and then generalize our 
results for arbitrary $m$ ($m=0,1,..$). The solutions for conventional GPT and $PT$ symmetric Scarf-II potentials 
are recovered for $m=0$. We further show the connection between the potential algebra approach
and SI for the extended cases.       

 The plan of the paper is as follows:
In Section 2, we briefly discuss the $SO(2,1)$ representation theory by modifying the generators
$J_{\pm}$ and write the expression for the potential and the corresponding
bound state energies in a closed form. In Section 3, 
we discuss two examples, one with real potential (extended GPT potential) and 
the other, a complex one (extended $PT$ symmetric Scarf-II). For the complex 
potential case, the corresponding potential algebra $sl(2,\mathbb{C})$
is also discussed in brief. In Section 4, we explain the shape invariance conditions in SQM and establish a connection between 
SI and potential algebra. Finally we summarize the results obtained and 
possible open questions in Section 5.

\section{The $SO(2,1)$ potential algebra and its realizations}

In this section, we construct the $SO(2,1)$ potential algebra 
 and its unitary representations. This algebra consists of three generators $J_{\pm}$ and $J_{3}$ and satisfy the commutation relations
\be\label{commut}
[J_{+},J_{-}]=-2J_{3};\qquad [J_{3},J_{\pm}]=\pm J_{\pm}\,.
\ee
The differential realization of these generators (corresponding to the 
well known solvable potentials) in $SO(2,1)$ algebra \cite{potalg8} is given by
\ba\label{udre}
J_{\pm}&=&e^{\pm i\phi}\bigg[\pm\frac{\partial}{\partial x}
-\bigg( (-i\frac{\partial}{\partial \phi}\pm\frac{1}{2})F(x)
-G(x)\bigg)\bigg]\,,\nonumber \\
J_{3}&=&-i\frac{\partial}{\partial\phi}\,.
\ea
However, we find that these generators are not sufficient to explain the 
spectrum of the extended SI potentials. Hence, 
we construct the $SO(2,1)$ algebra by modifying $J_{\pm}$ with the inclusion of a new operator, $U(x,-i\frac{\partial}{\partial \phi}\pm\frac{1}{2})$ 
as, 
\be\label{extdre}
J_{\pm}=e^{\pm i\phi}\bigg[\pm\frac{\partial}{\partial x}
-\bigg( (-i\frac{\partial}{\partial \phi}\pm\frac{1}{2})F(x)
-G(x)\bigg)-U(x,-i\frac{\partial}{\partial \phi}\pm\frac{1}{2})\bigg]\,,
\ee
and keeping the generator $J_{3}$ unchanged.\\
Here $F(x)$, $G(x)$ and $U(x,-i\frac{\partial}{\partial\phi}\pm\frac{1}{2})$ are two functions and a functional operator respectively. 
This functional operator $U(x,-i\frac{\partial}{\partial\phi}\pm\frac{1}{2})$
act on a basis \ket{k,m} to give a function $U(x,k\pm\frac{1}{2})$. These 
three functions will be different for the different potentials.

In order to satisfy the $SO(2,1)$ algebra (\ref{commut}) by these new generators $J_{\pm}$ and $J_{3}$, the following  
restrictions on the functions $F(x)$, $G(x)$ and $U(x, k\pm\frac{1}{2})$ 
\ba\label{rest1}
\frac{d}{dx}F(x)+F^2(x)=1;\qquad \frac{d}{dx}G(x)+F(x)G(x)=0;
\ea
and
\ba\label{rest2}
\bigg[U^2(x,k-\frac{1}{2})-\frac{d}{dx}U(x,k-\frac{1}{2})+2U(x,k-\frac{1}{2})\bigg( F(x)(k-\frac{1}{2})-G(x)\bigg)\bigg]-\nonumber\\
\bigg[U^2(x,k+\frac{1}{2})+\frac{d}{dx}U(x,k+\frac{1}{2})
+2U(x,k+\frac{1}{2})\bigg( F(x)(k+\frac{1}{2})-G(x)\bigg)\bigg]=0\,
\ea 
are required.

Note that Eq. (\ref{rest1}) is the same as for the usual potentials 
\cite{potalg8} while an additional condition (\ref{rest2}) appears
due to the presence of the extra term 
$U(x,-i\frac{\partial}{\partial \phi}\pm\frac{1}{2})$ in $J_{\pm}$.   
It may be noted that for this algebra the functions $F(x)$, $G(x)$ and $U(x,k\pm\frac{1}{2})$ are all real and the generators
$J_{+}$ and $J_{-}$ are Hermitian conjugate (i.e., $J_{+}$=$J^{\dagger}_{-}$) to each other. 
For a given values of $F(x)$, the function $G(x)$ is obtained by solving the first order linear differential Eq. (\ref{rest1})
and then the corresponding $U(x,k\pm\frac{1}{2})$ is obtained from Eq. (\ref{rest2}).

The Casimir operator, for the $SO(2,1)$ algebra, in terms of the above 
generators is given by
\be\label{cas}
J^2=J^2_{3}-\frac{1}{2}(J_{+}J_{-}+J_{-}J_{+}) 
= J^2_{3}\mp J_{3}-J_{\pm}J_{\mp}\,.
\ee 
For the bound states, the basis for an irreducible representation of 
extended $SO(2,1)$ is characterized by
\be
J^2\ket{j,k}=j(j+1)\ket{j,k};\qquad J_{3}\ket{j,k}=k\ket{j,k}\,,
\ee 
and
\be
J_{\pm}\ket{j,k}=[-(j\mp k)(j\pm k+1) ]^{\frac{1}{2}}\ket{j,k\pm1}\,.
\ee 
Using (\ref{extdre}), the differential realization of the Casimir operator 
in terms of $F(x)$, $G(x)$ and $U(x, J_{3}-\frac{1}{2})$ is given by 
\ba\label{dcasmir}
J^2&=&\frac{d^2}{dx^2}+\bigg(1-F^2(x)\bigg)(J^2_{3}-\frac{1}{4}) - 2\frac{dG(x)}{dx}(J_{3})-G^2-\frac{1}{4}\nonumber\\
&-&\bigg[U^2(x,J_{3}-\frac{1}{2})+\bigg(\big(J_{3}-\frac{1}{2}\big) F(x)-G(x)\bigg) U(x,J_{3}-\frac{1}{2})\nonumber\\
&+&U(x,J_{3}-\frac{1}{2})\bigg(\big( J_{3}
-\frac{1}{2}\big) F(x)-G(x)\bigg)-\frac{d}{dx}U(x,J_{3}-\frac{1}{2})\bigg]\,,
\ea
 and the basis \ket{j,k} in the form of function is given as
 \be\label{basis}
 \ket{j,k}=\psi_{jk}(x,\phi)\simeq \psi_{jk}(x)e^{ik\phi}\,.
 \ee
The functions (\ref{basis}) satisfy the Schro\"odinger equation 
\be\label{sch}
\bigg[-\frac{d^2}{dx^2}+V_{k}(x)\bigg]\psi_{jk}(x)=E\psi_{jk}(x)\,,
\ee
where $V_{k}(x)$ is one parameter family of $k$-dependent potentials given by
\ba\label{kpot}
V_{k}(x)&=&(F^2(x)-1)(k^2-\frac{1}{4})+2k\frac{d}{dx}G(x)+G^2(x)+(k-\frac{1}{2})^2\nonumber\\
&+&\bigg[U^2(x,k-\frac{1}{2})+2\bigg(\big( k-\frac{1}{2}\big) F(x)-G(x)\bigg) U(x,k-\frac{1}{2})\nonumber\\
&-&\frac{d}{dx}U(x,k-\frac{1}{2})\bigg]\,,
\ea
and the corresponding energy eigenvalues are given by
\be\label{eng}
E_{j}=\big(k-\frac{1}{2}\big)^2-\big(j+\frac{1}{2}\big)^2\,.
\ee
Thus the Hamiltonian in terms of the Casimir operator of $SO(2,1)$ algebra is 
given by
\be\label{ham}
H=-\big(J^2+\frac{1}{4}\big)\,.
\ee
\subsection{Unitary representation of $SO(2,1)$ algebra}

It may be noted that the $SO(2,1)$ algebra (\ref{extdre}) with the modified 
generators satisfies the same unitary representation as satisfied by the 
generators corresponding to the usual potentials \cite{potalg8}. Here we 
discuss the unitary representation of $SO(2,1)$ algebra corresponding to 
the discrete principal series $D^{+}_{j}$ for which $j<0$ i.e.,
\be\label{uniprep1}
j=-\frac{n}{2}-\frac{1}{2} \quad \mbox{or}\quad k=-j+n;\quad n=0,1,2,...,.  
\ee
Thus the energy eigenvalues (\ref{eng}) corresponding to this series will be
\be\label{eng1}
E_{n}=\big(k-\frac{1}{2}\big)^2-\big(n-(k-\frac{1}{2})\big)^2\,.
\ee
Once we fix the potential parameter $k$, then one has a finite number
of bound states ($n=0,1,2,...\{k-\frac{1}{2}\}$).

\section{Rationally extended potentials and its bound states}

In this section we consider two rationally extended SI potentials and obtain 
their solutions in  
terms of EOPs. We modify the generators $J_{\pm}$ by introducing $U(x,k\pm\frac{1}{2})$ 
appropriately for these potentials so that the condition (\ref{rest2}) is
satisfied and then obtain the exact bound state spectrum for both 
the cases.
 
\subsection{Rationally extended generalized P\"oschl-Teller (GPT) potential}

For this potential, we consider 
\be 
 F(x)=\coth x;\qquad G(x)=B\cosech x\,, \nonumber
\ee
and choose
\be
U(x,k-\frac{1}{2})=\bigg(\frac{2B\sinh x}{2B \cosh x-2(k-\frac{1}{2})-1}
-\frac{2B\sinh x}{2B \cosh x-2(k-\frac{1}{2})+1}\bigg); 
\quad B > k+\frac{1}{2} > 1\,,
\ee
so that the condition (\ref{rest2}) is satisfied.
On substituting these functions in (\ref{kpot}), we get the rationally 
extended GPT potential (which is defined on the half-line $0\leq x\leq\infty$) 
given by
\ba\label{gpt1}
V_{I}(x,k)= V_{GPT}(x,k)+ V_{rat}(x,k)\,,
\ea
where 
\be
V_{GPT}(x,k)= (k-\frac{1}{2})^2+[B^2 + (k-\frac{1}{2})(k+\frac{1}{2})] 
\cosech^2 x - B(2(k-\frac{1}{2}) + 1) \cosech x \coth x\,,
\ee
is the conventional GPT potential given in \cite{cks}, while
\be
V_{rat}(x,k)=\frac{4k}{(2B\cosh x-2k)}
-\frac{2[4B^2-(2k)^2]}{(2B\cosh x-2k)^2}\,,
\ee
is the rational part of the extended potential $V_{I}(x,k)$.
The energy eigenvalues of this extended potential are same as that of 
conventional one (i.e they are isospectral) and are given by 
Eq. (\ref{eng1}) i.e.
\be\label{engg}
E_n = (k-\frac{1}{2})-(n-(k-\frac{1}{2}))^2;\quad n= 0,1, . . . ,n_{max};
\quad (k-\frac{3}{2})\leq n_{max}<(k-\frac{1}{2})\,.
\ee
This extended GPT potential is same as given in Ref. \cite{bqr,qscat1} with 
parameter $A$ replaced by $(k-\frac{1}{2})$ and the  
associated wavefunctions $\psi_{jk}(x)$ (\ref{sch}) are given in terms of $X_1$ exceptional Jacobi polynomial.

The potentials corresponding to $X_{m}$ exceptional Jacobi polynomials can be obtained by considering 
\ba
U(x,k\pm\frac{1}{2})&\Rightarrow &U(x,m,k\pm\frac{1}{2}) \nonumber \\
&=& \frac{(m-2B-1)\sinh x}{2}\times\bigg[\frac{P^{(-B+(k\pm\frac{1}{2})+\frac{1}{2},-B-(k\pm\frac{1}{2})-\frac{1}{2})}_{m-1}(\cosh x)}{P^{(-B+(k\pm\frac{1}{2})-\frac{1}{2},-B-(k\pm\frac{1}{2})-\frac{3}{2})}_{m}(\cosh x)}\nonumber\\
&-&\frac{P^{(-B+(k\pm\frac{1}{2})-\frac{1}{2},-B-(k\pm\frac{1}{2})
+\frac{1}{2})}_{m-1}(\cosh x)}{P^{(-B+(k\pm\frac{1}{2})-\frac{3}{2},
-B-(k\pm\frac{1}{2})-\frac{1}{2})}_{m}(\cosh x)}\bigg]\,,
\ea
where $P^{(\alpha, \beta)}_{m}(\cosh x)$ with $\alpha=B-(k\pm\frac{1}{2})
-\frac{1}{2}$ and $\beta=-B-(k\pm\frac{1}{2})-\frac{1}{2}$, is conventional
Jacobi polynomial. The energy eigenvalues will be same (since the spectrum are isospectral to the usual one) as given in Eq. (\ref{engg}).

For $m=0$, the function $U(x,m,k\pm\frac{1}{2})$ becomes zero, hence we obtain the usual $SO(2,1)$ algebra as satisfied by Eq. (\ref{udre}) and (\ref{rest1}),
and the corresponding potential will be the usual GPT potential. However for 
$m=1$, we recover the results obtained above corresponding to the $X_{1}$ 
exceptional polynomial case.

\subsection{Rationally extended $PT$ symmetric complex Scarf-II potential}

In this section we consider the rationally extended $PT$ symmetric complex 
Scarf-II potential. Note that the 
real Scarf-II potential can not be extended rationally due to the presence of 
the singularity in the wavefunction \cite{bqr}.
For this complex potential we use extended $sl(2,\mathbb{C})$ potential algebra to find its solution in terms of EOPs.

 In this approach at least one of functions  $F(x)$, $G(x)$ and 
$U(x, k\pm\frac{1}{2})$ must be complex  
and satisfy Eqs. (\ref{rest1}) and (\ref{rest2}). As a result, unlike the 
earlier case, the generators defined 
in (\ref{extdre}) for this 
potential are not Hermitian conjugate of each other (i.e, 
$J_{-}\neq J^{\dagger}_{+}$). 

To solve the extended $PT$ symmetric Scarf-II potential, we consider the 
functions
\be
F(x)=\tanh x;\qquad G(x)=iB \sech x\,,
\ee
and construct 
\be
U(x,k\pm\frac{1}{2})= \bigg[ \frac{2iB\cosh x}{(-2iB\sinh x+2(k\pm\frac{1}{2})
-1)}-\frac{2iB\cosh x}{(-2iB\sinh x+2(k\pm\frac{1}{2})+1)}\bigg]\,,
\ee 
such that Eqs. (\ref{rest1}) and (\ref{rest2}) are satisfied. \\
Substituting all these functions in (\ref{kpot}), we get the rationally extended $PT$ symmetric potential (which is on the full-line $-\infty\leq x\leq\infty$) given by
\ba\label{scarf}
V_{II}(x,k)= V_{Scarf}(x,k)+ V_{rat}(x,k)\,,
\ea
where 
\be
V_{Scarf}(x,k)= (k-\frac{1}{2})^2+[(iB)^2 - (k-\frac{1}{2})(k+\frac{1}{2})] 
\sech^2 x + iB(2(k-\frac{1}{2}) + 1) \sech x \tanh x\,,
\ee
is the conventional $PT$ symmetric Scarf-II potential \cite{pt1} withe the 
parameter $A$ being replaced by $(k-\frac{1}{2})$ and
\be
V_{rat}(x,k)=\frac{-4k}{(-2iB\sinh x+2k)}+\frac{2[4(iB)^2
+(2k)^2]}{(-2iB\sinh x+2k)^2}\,,
\ee
is the rational part of the extended potential.
The energy eigenvalues for this extended complex potential are real and are
the same as that of the conventional one and are given by
\be\label{enscarf} 
E_n = (k-\frac{1}{2})-(n-(k-\frac{1}{2}))^2;\quad n= 0,1, . . . ,n_{max};
\qquad n_{max} < (k+\frac{1}{2})\,.
\ee
This extended $PT$ symmetric Scraf-II potential is the same as given in 
\cite{bqr} with parameter $A$ being replaced by $(k-\frac{1}{2})$ and the  
associated wavefunctions $\psi_{jk}(x)$ (\ref{sch}) are given in terms of $X_1$ exceptional Jacobi polynomial.
 
Similar to the extended GPT case, the extended $PT$ symmetric Scarf-II 
potentials associated with the $X_m$ exceptional polynomials 
are again obtained by modifying the function, i.e.
\ba
U(x,k\pm\frac{1}{2})&\Rightarrow &U(x,m,k\pm\frac{1}{2}) \nonumber \\
&=& \frac{(m-2B-1)i\cosh x}{2}\times\bigg[\frac{P^{(-B+(k\pm\frac{1}{2})+\frac{1}{2},-B-(k\pm\frac{1}{2})-\frac{1}{2})}_{m-1}(i\sinh x)}{P^{(-B+(k\pm\frac{1}{2})-\frac{1}{2},-B-(k\pm\frac{1}{2})-\frac{3}{2})}_{m}(i\sinh x)}\nonumber\\
&-&\frac{P^{(-B+(k\pm\frac{1}{2})-\frac{1}{2},-B-(k\pm\frac{1}{2})
+\frac{1}{2})}_{m-1}(i\sinh x)}{P^{(-B+(k\pm\frac{1}{2})
-\frac{3}{2},-B-(k\pm\frac{1}{2})-\frac{1}{2})}_{m}(i\sinh x)}\bigg]\,,
\ea
where $P^{(\alpha, \beta)}_{m}(i\sinh x)$ with $\alpha=B-(k\pm\frac{1}{2})
-\frac{1}{2}$ and $\beta=-B-(k\pm\frac{1}{2})-\frac{1}{2}$, is conventional
Jacobi polynomial. The energy eigenvalues will be the same (since the spectrum 
is isospectral to the usual one) and is given in Eq. (\ref{enscarf}).

For $m=0$, the function $U(x,m,k\pm\frac{1}{2})$ becomes zero, hence we obtain the usual case of $sl(2,\mathbb{C})$ and the corresponding 
potential will be the usual $PT$ symmetric Scarf-II potential \cite{pt1}. On
the other hand, for $m=1$, we recover our results corresponding to the 
$X_{1}$ exceptional polynomials as discussed above.         
 
\section{Shape invariance and connection to extended Potential Algebra}

The connection between the shape invariance (SI) condition in SQM and the usual potential algebra 
was elegantly established in \cite{sipotalg1}. These authors showed that both 
the approaches are equivalent. In this section we  
show that the above connection continues to hold good even for the extended 
generators introduced by us.
Before showing this, first we briefly discuss about the shape invariance (SI) 
condition in SQM.    
A detailed description can be found, say for example in \cite{cks}.

If the supersymmetric partner potentials $V_1,_2(x)$ are
similar in shape and differ only in the values of the parameters, then they 
are said to be shape invariant and satisfy the condition
\be\label{si} 
V_2(x,a_0) = V_1(x,a_1)+R(a_0)\,,
\ee 
where $a_1$ is a function of $a_0$ and the remainder $R(a_0)$ is related to
the  ground state energy
of $V_2(x,a_0)$, because the ground state energy of $V_1(x,a_1)$ is zero by
construction. In the special case of SIP with translation, $a_1$ and $a_0$
differ by a constant.

For the superpotential $ W(x,k)$, the SI condition with translation implies 
\be
W^{2}(x,k)+ W^{'}(x,k) = W^{2}(x,k+1)- W^{'}(x,k+1)+ R(k)\,,
\ee
where $V_{1,2}(x,k)=W^2(x,k)\pm W'(x,k)$.\\
As we know, this constraint suffices to determine the entire spectrum of the potential $V_1(x,k)$.
 Since for SI potentials, the parameter $k$ is changed by a constant amount each time as one goes from 
the potential $V_1(x,k)$ to its superpartner, it is natural to ask whether such a task can be formally
accomplished by the action of a ladder type operator.

With that in mind, in the extended potential algebra, if we redefine the 
operators $J_{\pm}$ (\ref{extdre}) in the form of an operator
$W(x,m,-i\frac{\partial}{\partial \phi}\pm\frac{1}{2})$ i.e.,
\be
J^{\pm }= e^{\pm i\phi} [\pm \frac{\partial }{\partial x}
- W(x,-i\frac{\partial }{\partial \phi}\pm \frac{1}{2})]\,,
\ee
where 
\be
W(x,-i\frac{\partial }{\partial \phi}\pm \frac{1}{2})=
W_1(x,-i\frac{\partial }{\partial \phi}\pm \frac{1}{2})
+W_2(x,-i\frac{\partial }{\partial \phi}\pm \frac{1}{2})\,,
\ee
is the operator form of the superpotential. \\
In terms of $F(x)$, $G(x)$ and $U(x,-i\frac{\partial}{\partial\phi}\pm\frac{1}{2})$, the above operators are give by
\be
W_1(x,-i\frac{\partial }{\partial \phi}\pm \frac{1}{2})=\bigg((-i\frac{\partial}{\partial \phi}\pm\frac{1}{2})F(x)-G(x)\bigg)
\ee
and
\be
W_2(x,-i\frac{\partial }{\partial \phi}\pm \frac{1}{2})=U(x,-i\frac{\partial}{\partial \phi}\pm\frac{1}{2}).
\ee 
Using these operators the commutation relation $[J_{+},J_{-}]$ is given by
\ba
[J_{+},J_{-}]&=&\bigg[-\frac{\partial ^{2}}{\partial x^{2}} + W^{2}(x,J_3-\frac{1}{2})- W{'}(x,J_3-\frac{1}{2})\bigg]\nonumber\\
&-&\bigg[-\frac{\partial ^{2}}{\partial x^{2}} + W^{2}(x,J_3+\frac{1}{2})+ W{'}(x,J_3+\frac{1}{2})\bigg].
\ea
 Using the SI condition i.e. 
\be
V_1(x,J_3-\frac{1}{2})- V_2(x,J_3+\frac{1}{2}) = -R(J_3+\frac{1}{2})\,, 
\ee
we get
\be
 [J^{+},J^{-}]=R(J_3+ \frac{1}{2}).
\ee
One can also explicitly check that the second commutation relation 
$[J_3,J_{\pm}]$=$\pm J_{\pm}$ is indeed satisfied.\\
Thus we see that similar to the usual one, in extended case also, the 
SI enables us to close the algebra 
of $J_3$ and $J_{\pm}$ to
\be\label{sialg}
 [J_3,J_{\pm }]= \pm J_{\pm };\qquad [J_{+},J_{-}]= -R\big(J_3+\frac{1}{2}\big).
 \ee  
If the function $R(J_3)$ is linear in $J_3$, then the above 
algebra (\ref{sialg}) would
reduce to that of a $SO(2,1)$ or $sl(2,\mathbb{C})$.  Both of the extended (translationally) shape 
invariant potentials (i.e., extended GPT and $PT$ symmetric 
Scarf-II) satisfy these conditions. For these two potentials the function $R\big(J_3+\frac{1}{2}\big)$ reduces to
$2J_3$ and the Eq. (\ref{sialg}) reduces to an $SO(2,1)$ or $sl(2,\mathbb{C})$ potential algebra, and thus establishes
the connection between SI and the potential algebra. The same can also be 
established easily for the $X_m$ case.  
 
\section{Summary and discussion}
In this work we have modified the generators of $SO(2,1)$ and $sl(2,\mathbb{C})$ in such a manner 
that they still satisfy the algebra of the corresponding groups subjected to certain conditions.
Further we have shown that the Hamiltonian for the rationally extended GPT 
and $PT$ symmetric  Scarf-II systems are expressed purely in terms of the 
modified Casimir operator of $SO(2,1)$ and $sl(2,\mathbb{C})$ 
groups respectively. This important realization enable us to obtain the spectrum of these rationally 
extended systems, whose solutions are in terms of EOPs in a closed form. We 
reproduce the 
solutions of GPT and $PT$ symmetric Scarf-II potential as a limiting case ($m=0$) of our results.
A connection between these algebras and the shape invariance (with translation) 
of SQM has also been established. 

It may be noted that in this paper we have only worked with the extended 
potentials which are translationally shape invariant and for which
the function $R(J_{3}+\frac{1}{2})$ is a linear function of $J_3$ only. There 
are other rationally
extended potentials \cite{op4} for which the usual shape invariance condition 
is not valid and instead they
satisfy an unfamiliar extended SI condition. 
So it will be interesting to know whether there are potential algebra
that describe these systems. Currently these problems are under investigation.\\

{\bf Acknowledgments}

R.K.Y acknowledges S. K. M. University for providing study leave and B.P.M acknowledges the financial support 
from the Department of Science and Technology (DST), Gov. of India under SERC project sanction
grant No. $SR/S2/HEP-0009/2012$. AK acknowledges financial support from Dept. 
of Atomic Energy in the form of Raja Ramanna Fellowship.

\end{document}